%% file: main.tex
\documentclass[conference]{IEEEtran}
\IEEEoverridecommandlockouts
\usepackage[utf8]{inputenc}
\usepackage{cite}
\usepackage{amsmath,amssymb,amsfonts}
\usepackage{graphicx}
\usepackage{textcomp}
\usepackage{xcolor}
\usepackage{amssymb}
\def\BibTeX{{\rm B\kern-.05em{\sc i\kern-.025em b}\kern-.08em
    T\kern-.1667em\lower.7ex\hbox{E}\kern-.125emX}}

\usepackage{enumitem}
\input{macros}
\input{macros2}

\begin{document}

\newcommand{\attackname}{{MIND-Crypt}}

\newcommand{\jimmy}[1]{%
  \textcolor{red}{\textbf{Jimmy's comment:}} \textcolor{red}{#1}%
}

\title{Robust and Verifiable MPC with Applications \\to Linear Machine Learning Inference}



\author{%
  \IEEEauthorblockN{Tzu-Shen Wang\IEEEauthorrefmark{1},
                    Jimmy Dani\IEEEauthorrefmark{1},
                    Juan Garay\IEEEauthorrefmark{1},
                    Soamar Homsi\IEEEauthorrefmark{2},
                    Nitesh Saxena\IEEEauthorrefmark{1}}\\
  \IEEEauthorblockA{\IEEEauthorrefmark{1}\textit{Department of Computer Science}, Texas A\&M University, College Station, TX, USA}\\
  \IEEEauthorblockA{\IEEEauthorrefmark{2}\textit{Air Force Research Laboratory}, Wright-Patterson AFB, OH, USA}\\
  \IEEEauthorblockA{Emails: \{jasonwang017, danijy, garay, nsaxena\}@tamu.edu, soamar.homsi@us.af.mil}%
}


\pagestyle{plain}

\maketitle

\input{abstract}


\input{introduction}
\input{preliminaries}
\input{robust-and-verifiable-mpc}
\input{applications-and-experiment-results}



\bibliographystyle{ieeetr}
\bibliography{references}

\end{document}

%% file: macros2.tex
\newcommand{\func}[1][\relax]{\ensuremath{\mathcal{F}_{\mathsf{#1}}}\xspace}

\newcommand{\fcaba}
{\func[conc\mhyphen a\mhyphen ba]}

\newcommand{\Comm}{\textrm{Comm}}
\newcommand{\Enc}{\textrm{Enc}}

\ifdefined\IsLLNCS

\else

\fi

\mathchardef\mhyphen="2D

\newcommand \mycaption {\small }     
\newcommand \mylabel {}

\ifdefined\IsLLNCS
    \newenvironment{nfbox}[3]{
    \renewcommand \mycaption {#1}
    \renewcommand \mylabel {#2}
    \begin{center}\small
    \begin{tabular}{|ll|}
    \hline
    \hspace{.3ex}
    \begin{minipage}{.97\linewidth}
         \vspace{0.5ex}
         #3}
         {\smallskip
     \end{minipage}
     &\hspace{.3ex} \\
     \hline
     \end{tabular}
     \captionof{figure}{\mycaption}
     \label{\mylabel}
     \end{center}    
    }
\else
    
\fi

\ifdefined\IsLLNCS

\spnewtheorem{vasclaim}[theorem]{Claim}{\itshape}{\normalfont}

\else

\newcounter{thm}
\newcommand{\funcbox}[3][Functionality]{
    \fbox{%
        \begin{minipage}{0.45\textwidth}
        \footnotesize
        \begin{center}
            {\textbf{#1} \ensuremath{#2}} 
        \end{center}
        \hspace{10pt}
        \\
        #3
        \end{minipage}%
    }%
}

\newcommand{\protbox}[2]{\funcbox[Protocol]{#1}{#2}}

\newcommand{\simbox}[2]{\funcbox[Simulator]{#1}{#2}}

%% file: abstract.tex
\begin{abstract}
In this work, we present an efficient secure multi-party computation MPC protocol that provides strong security guarantees in settings with a dishonest majority of participants who may behave arbitrarily. 
Unlike the popular MPC implementation known as {\em SPDZ} [Crypto '12],
which only ensures security 
with abort, our protocol achieves both {\em complete identifiability} and {\em robustness}. With complete identifiability, honest parties can detect and unanimously agree on the identity of any malicious party. Robustness allows the protocol to continue with the computation without requiring a restart, even when malicious behavior is detected. Additionally, our approach addresses the performance limitations observed in the protocol by Cunningham {\em et al.} [ICITS '17],
which, while achieving complete identifiability, is hindered by the costly exponentiation operations required by 
the choice of commitment scheme.

Our protocol is based on the approach by Rivinius {\em et al.} [S\&P '22], utilizing lattice-based commitment for better efficiency.
We
achieves robustness with the help of a semi-honest trusted third party. 
We benchmark our robust protocol, 
showing the efficient recovery 
from 
parties' malicious behavior.  

Finally, we benchmark our protocol on a ML-as-a-service scenario,
wherein clients off-load the desired computation to the servers, 
and verify the computation result. We benchmark on linear ML inference, running on various datasets. While our efficiency is slightly lower compared to SPDZ's, we offer stronger security properties that provide distinct advantages.
\end{abstract}

%% file: introduction.tex
\section{Introduction}

Outsourcing computation to cloud servers has become an invaluable practice in today’s digital landscape. By off-loading intensive computational tasks to remote data centers, clients gain a cost-effective solution that eliminates the need for significant investment in hardware and software infrastructure. Instead, they can leverage the vast, on-demand computing power of the cloud to scale resources as needed. This flexibility not only reduces initial capital expenses but also enhances operational efficiency, enabling organizations to focus on core activities and innovation rather than managing complex IT systems. In an age defined by data-driven decisions and optimized resource use, cloud outsourcing is now essential for businesses and individuals alike.

While outsourcing to the cloud offers numerous advantages, it also introduces significant security challenges. Cloud-hosted data and applications are susceptible to risks such as data breaches, unauthorized access, and service outages. Clients must depend on the cloud provider's security protocols, which may not always align with their unique requirements and standards. Ensuring data privacy, regulatory compliance, and control over sensitive information becomes more complex when data resides on remote servers. In this work, we focus on safeguarding client data privacy and ensuring the accuracy of outsourced computations.

Secure multi-party computation (MPC)~\cite{DBLP:conf/focs/Yao82,GoldreichMW87,BGW88,CCD88} is a powerful tool for enhancing the security of outsourced cloud computing. MPC enables multiple parties to jointly compute a function over their inputs while preserving the privacy of those inputs, ensuring data confidentiality within a cloud environment. This approach allows clients to securely delegate computational tasks to the cloud, protecting sensitive information from potential exposure, even if the cloud service provider is untrusted. As a cryptographic technique, MPC is essential for mitigating the security risks associated with cloud outsourcing, by utilizing multiple service providers. However, existing MPC protocols have certain drawbacks—some lack efficiency, while others fall short in providing robust security guarantees.

In this paper we are interested in 
guaranteeing security even in the presence of a dishonest majority of service providers.
In such a setting --- MPC with a dishonest majority --- protocols can be categorized into two main types:  
\begin{tiret}
    \item Protocols with common security guarantees, exemplified by efficient and widely deployed solutions like SPDZ~\cite{DBLP:conf/crypto/DamgardPSZ12}. These protocols offer security with abort, meaning that if misbehavior by a party is detected, the protocol execution is aborted. As a result, the computation may fail to complete successfully.
    \item Stronger security guarantees, such as {\em robustness}, which guarantees that malicious parties cannot prevent the honest parties from obtaining the output of the computation, as well as {\em (complete) identification} of the misbehaving parties. As shown in~\cite{DBLP:conf/icits/CunninghamFY17}, the latter can be achieved at the expense of efficiency, or by more sophisticated, lattice-based cryptography methods~\cite{DBLP:conf/sp/RiviniusR0K22}. 
\end{tiret}
\ignore{As the protocol in \cite{DBLP:conf/sp/RiviniusR0K22} 
utilizes a lattice-based commitment scheme, 
it also provides complete identifiability but 
much better efficiency compared with the~\cite{DBLP:conf/icits/CunninghamFY17} protocol.
} 
\subsection{Our Contributions}\label{subsec:contributions}
In this work we 
enhance the approach in~\cite{DBLP:conf/sp/RiviniusR0K22} 
by introducing an additional entity, namely, a {\em semi-honest trusted third party} (STTP, which can be one of the clients)
to 
achieve robustness 
for a number of corruptions of up to $n-2$ parties. (For the reason for dissimilar thresholds --- i.e., $n-2$ vs. $n-1$ --- with a trusted dealer, we 
can identify every malicious server and open its share to the other servers. However, if we identify $n-1$ malicious servers and open their shares, it would lead to the only remaining server being able to combine the $n-1$ servers' shares and its own share to recover the input, which is a situation we wish to avoid; therefore, we ``degrade'' our guarantees to security with abort) 

In~\cite{DBLP:conf/sp/RiviniusR0K22}, 
a tradeoff is made between privacy and robustness. With threshold $t$ used to reconstruct shares, their protocol fails to provide robustness if there are more than $n-t$ malicious parties, and fails to provide privacy if there are more than $t$ malicious parties.  In contrast, our protocol achieves privacy if there is at least one honest party, and achieves robustness when there are less than $n-2$ malicious parties.
Moreover, our 
protocol does not need to restart as malicious behavior is detected,
\ignore{The scenario we are aiming for is MPC for a service, the client wants to delegate their computation to the MPC server, while still protecting the client's private data. On the server side, the servers also want to be financially efficient, therefore, they may throw away the secret share they receive from the client. Therefore, this does not allow the server just to restart the protocol from the bottom of the computation circuit.
} 
which we achieve by means of
homomorphic encryption.

We showcase the performance of our protocol by benchmarking it
on neural network Network-A \cite{DBLP:conf/sp/MohasselZ17,DBLP:conf/ccs/Keller20,DBLP:conf/sp/RiviniusR0K22}, which 
consists of a sequence of {\em dense}
and {\em square} layers. In more detail, a {\em neuron} in
the dense layer includes a weighted sum of all previous layers (or the input in case of the first layer), and  it captures how much influence of each value from previous layers should be considered. 
On the other hand, the square layer adds non-linearity to the output of the dense layer; it provides features such as avoiding over-fitting and capturing more relationships that cannot be explained by using linear relations on the inputs. 
Further, we also benchmark our protocol on a linear ML application, where we show that accuracy is not lost, 
while achieving reasonable efficiency. 

In ML-as-a-service scenarios, clients secret-share their inputs with the computation servers.
Once the computation is complete, the servers return the output to the client.
Our MPC protocol, 
is ``batch-based,'' in the sense that it achieves {\em amortized} efficiency
by running computations on collections (batches) of input data; as such
it is well-suited for ML-as-a-service applications, where, in order to optimize hardware utilization and reduce costs, the computation servers may prefer processing client requests in batches rather than individually. 
Since our MPC protocol operates over polynomial rings, batch processing of multiple inputs is inherently enabled, as these rings can be decomposed into multiple slots, with each slot encoding an input (cf.~\cite{DBLP:journals/iacr/GentryHS12}). 


\ignore{
\cite{DBLP:conf/sp/RiviniusR0K22} makes use of a lattice-based commitments. And in \cite{DBLP:conf/sp/RiviniusR0K22} shows how such commitments can be homomorphically updated, so that they stay consistent with the new messages being exchanged during CESS's execution.
The online protocol of \cite{DBLP:conf/sp/RiviniusR0K22} utilizes lattice based commitments to authenticate messages, in particular when a server $S_i$ opens a message to other servers $S$/$S_i$. If $S_i$ cheats, its misbehavior is identified. 
} 

\subsection{Related Work}
Our work ensures public verifiability, complete identifiability, and robustness in the presence of a dishonest majority.
Related works along the SPDZ line of work
(e.g.,~\cite{DBLP:conf/crypto/DamgardPSZ12,DBLP:conf/ccs/Keller20,DBLP:journals/iacr/KellerPR17,DBLP:conf/ccs/KellerOS16}), 
improve the efficiency of the online computation phase, 
while still only achieving 
security with abort.

Other related work, such as \cite{DBLP:journals/iacr/Scholl0S21} also utilize a bulletin board, enabling public verifiability. Third parties use the information published on the bulletin board with the messages opened during the computation to verify the correctness of the computation.

Regarding 
security with identifiable abort, there are also works  
that enable honest parties to detect malicious behavior and identify the corresponding parties, such as~\cite{DBLP:conf/sp/RiviniusR0K22}. 
The protocol in~\cite{DBLP:conf/sp/RiviniusR0K22}, however, only provide robustness when there is an honest majority; 
otherwise, privacy will be violated. In contrast, by adding an STTP, our protocol provides robustness even under a dishonest majority and enables honest parties to recover shares held by the malicious party without having to restart the protocol.

In addition, the presence of an STTP
allows us to achieve fairness even with a dishonest majority. Specifically, if a malicious party refuses to open its secret share, the STTP and an arbitrary honest party can pool their shares
and reconstruct it.
Thus, a dishonest party cannot abort with an advantage. 
In contrast, if in \cite{DBLP:conf/sp/RiviniusR0K22} there is a dishonest majority, those parties can learn the secret share themselves and from that point on refuse to participate in the protocol.

Similarly to ~\cite{DBLP:conf/sp/RiviniusR0K22}, 
our approach applies to amortized settings, where multiple requests are served in tandem. 
In addition to providing
better hardware 
utilization, the approach is a suitable candidate for MPC-as-a-service, as argued above.
\subsection{Organization of the Paper}
The organization of the rest of the paper is as follows. 
Section~\ref{sec:prelim} describes the network and computational model and
lists 
the building blocks that are used by our construction, including but not limited to 
lattice-based commitments, homomorphic encryption, and distributed decryption. Our robust and verifiable MPC protocol is presented in detail in Section~\ref{sec:protocol}, together with its security analysis. 
Section~\ref{sec:experiments} 
focuses on experimental results: The benchmarking of Network A appears in Section~\ref{subsec:network-A}, while the benchmarking
of linear ML computations appears in Section~\ref{subsec:ml-framework-design}. 

%% file: preliminaries.tex
\section{Preliminaries}
\label{sec:prelim}
\subsection{System Model}\label{sec:model}
As it is customary, we model protocol participants as probabilistic polynomial-time Turing machines (ITMs) 
and consider the client-server model of computation with an STTP. We assume a point-to-point synchronous communication network, a public-key infrastructure (PKI), and 
a bulletin board (for simplicity, as it can be realized 
from the PKI). Table~\ref{tab:notation} summarizes the notation used in our protocol descriptions. 
\begin{table} [h!]
  \centering
  \caption{Summary of notation used in the paper.}
  \label{tab:notation}
  \begin{tabular}{|p{2cm}|p{6cm}|}
    \hline
   Symbols & Definition. \\
    \hline
  ${\cal C}$ & A set of clients $\{C_1, ..., C_m\}$\\
    \hline
    ${\mathcal S}$ & A set of servers $\{S_1, ...,S_n\}$\\
    \hline
    STTP & Semi-honest trusted third party\\
    \hline
    ${\cal B}$ & Bulletin board (broadcast channel) \\
    \hline
    $[x]$ & Secret share of value $x$ (e.g., a client's input) 
    \\
    \hline
    ${\cal P}$ & Prover in ZK proof (e.g., $\Sigma$ protocol) \\ 
    \hline
    ${\cal V}$ & Verifier in ZK proof 
    \\
    \hline
  \end{tabular}
\end{table}

\ignore{As mentioned above, we work in the client/server model where a set of clients ${\cal C}$ = $\{C_1, ..., C_m\}$ wish to offload an arbitrary computation on their inputs to a set of servers ${\cal S} = \{S_1, ...,S_n\}$. In addition, we assume the existence of a {\em semi-honest trusted third party} (STTP), which in fact could be one of the clients \Soamar{Third party cannot be one of the client. We already discussed this possibility }, at the expense of having to monitor the execution of the protocol. Our network model assumes the existence of a broadcast channel, or ``bulletin board,'' denoted by $\cal B$. \Soamar{provide a figure that includes all the components of the system model}

Our protocol will be making use of protocols instantiating commitment schemes, in which case we will be using $\cal S$ and $\cal R$ to denote the committer (the sender of the commitment) and the receiver, respectively, and protocols for ZK proofs, such as $\Sigma$-protocols~\cite{DBLP:conf/crypto/CramerDS94}, in which case will be using $\cal P$ and $\cal V$ to denote the prover and the verifier, respectively. We will use $[x]$ to denote the secret share of a value (e.g., a client's input) $x$,
} 
\subsection{Building Blocks}
\label{sec:building-blocks}
\paragraph{MPC.}
In secure multi-party computation (MPC)
\cite{Yao86,GoldreichMW87,BGW88}, $n$ parties hold 
input $x_1,...,x_n$ respectively, aiming to compute a given function $f(x_{1},...,x_{n})$ privately and correctly.
Below we list 
the basic security properties for MPC. 
\begin{tiret}
    \item {\em Privacy:}
    The 
    parties' inputs remain private.
    \item {\em Security with abort:}
    All honest parties 
    agree on abort.
    \item {\em Robustness:} The protocol always outputs a correct result regardless of the adversary ${\cal A}$'s behavior (also called {\em guaranteed output delivery}) (cf.~\cite{10.1145/73007.73014,10.5555/1756123.1756154}).
    \item {\em Complete identifiability:} When a corrupted party misbehaves,
    honest parties 
    always identify and agree on the identities of the misbehaving party
\end{tiret}
\paragraph{Commitments.} 
We define the commitment operation as
$\Comm(x, r)$, where committer commits to a message $x$ 
where $r$ is the randomness used in the commitment (r also acts as part of the decommitment in the opening phase).
The interface for the verification operation is given by $Ver(\Comm, x, r)$, where the verifier takes a commitment and checks if it is consistent with the committed message $x$ and the decommitment $r$.
The two basic properties of a commitment scheme are as follows (c.f. \cite{DBLP:books/crc/KatzLindell2007}):

\begin{tiret}
\item {\bf Hiding:} $\Comm(x,r)$ leaks no non-trival info of $x$. An adversary $\cal A$ breaks hiding iff with non\text{-}negl probability 
\begin{enumerate}
    \item Parameters $ \text{params} \leftarrow \text{Gen}(1^n) $ are generated.
    \item The adversary $ A $ is given input $ \text{params} $, and outputs a pair of messages $ m_0, m_1 \in \{0, 1\}^n $.
    \item A uniform $ b \in \{0, 1\} $ is chosen, and $ \text{com} \leftarrow \text{Com}(m_b, r) $ is computed.
    \item The adversary $ A $ is given $ \text{com} $ and outputs a bit $ b' $.
    \item The output of the experiment is 1 if and only if $ b' = b $.
\end{enumerate}
\item {\bf Binding:} An adversary $\cal A$ cannot open $\Comm(x,r)$ to $x{'}$, except with negligible probability. $\cal A$ breaks binding iff with non\text{-}negl probability
\begin{enumerate}
    \item Parameters $ \text{params} \leftarrow \text{Gen}(1^n) $ are generated.
    \item $ A $ is given input $ \text{params} $ and outputs $(comm, m, r, m_0, r_0)$.
    \item The output of the experiment is defined to be 1 if and only if $ m \neq m_0 $ and \\
    $ \Comm(m, r) = comm = \Comm(m_0, r_0). $
\end{enumerate}
\end{tiret}

In order to provide complete identifiability in our MPC scheme, 
committed values need to be 
updated as the computation proceeds.
The opening server computes its share x to $x^{'}$ and opens it to the receiving server. With the homomorphic property, the receiving server updates the commitment $\Comm(x)$ to $\Comm(x^{'})$, then uses $\Comm(x^{'})$ to authenticate $x^{'}$. By the binding property of the commitment, the authentication succeeds if and only if $x^{'}$ is correct.
As such, we will require 
the commitment scheme to be linearly homomorphically updateable, satisfying the following properties:


\begin{tiret}
\item $\Comm(x_1, r_1) + \Comm(x_2, r_2) = \Comm(x_1+x_2, r_1+r_2)$
\item $\Comm(x_1, r_1) + c = \Comm(x_1+c,  r_1)$ 
\item $\Comm(x_1, r_1)*c = \Comm(cx_1, cr_1)$
\end{tiret}

Further, for efficiency reasons, we will be employing lattice-based commitments~\cite{DBLP:conf/sp/RiviniusR0K22} 
, which satisfy our homomorphic updates requirement.
In addition, such commitments can (and will) be used  
to authenticate messages, in particular during the course of the computation server $S_i$ opens a message to other servers $S \backslash S_i$; if $S_i$ cheats, its misbehavior is identified. With the binding property, the cheating server can not be opened to a different message without getting detected. 

With the homomorphic property, a server $S_i$ can update a commitment locally to any layer of the computation circuit. Thus, when another server intends to open its share, the server $S_i$ holds the commitment at the same circuit layer as the layer of opening. 
Further, due to the binding property,
the decommitment verification passes if and only if the opened share is correct. 
Lattice-based commitments require only simple operations, such as multiplication and addition, whereas discrete log-based commitments involve costly exponentiation. 

\paragraph{$\Sigma$ protocols.}
 This building block, proposed by Cramer {\em et al.}~\cite{DBLP:conf/crypto/CramerDS94}, can be used to provide a 
 ZK proof that both a given encryption and a Pedersen commitment correspond to the same value, say, $x$, 
 without revealing $x$.
We remark that 
such functionality can be converted into a non-interactive form using the Fiat-Shamir heuristic. 

$\Sigma$-protocols can be 
realized 
based on lattices
\cite{DBLP:conf/asiacrypt/Lyubashevsky09,DBLP:conf/eurocrypt/Lyubashevsky12}. 
Please refer to those papers of $\Sigma$ protocol for lattice-based signatures 
(with a similar approach for commitments) and it achieves non-interactivity via the Fiat-Shamir heuristic (cf.~\cite{DBLP:conf/asiacrypt/Lyubashevsky09}).
\cite{DBLP:conf/sp/RiviniusR0K22} shows that simulation proof can be constructed by allowing the simulator to generate a ``fake'' ZK
proof, assuming a programmable random oracle (RO).


\paragraph{Homomorphic encryption.}
To provide the robustness property in our MPC scheme, we require encryption to be homomorphic, so that the encryption can be updated along with the computation of the circuit: 

\begin{tiret}
\item $\Enc(x_1) + \Enc(x_2) = \Enc(x_1+x_2) $
\item $\Enc(x_1) + c = \Enc(x_1+c)$ 
\item $\Enc(x_1) \cdot c = \Enc(c x_1)$ 
\end{tiret}
\paragraph{BGV encryption.} 
BGV encryption~\cite{DBLP:journals/toct/BrakerskiGV14} is a fully homomorphic encryption scheme 
We also utilize distributed decryption from \cite{DBLP:conf/sp/RiviniusR0K22}.

\ignore{
\paragraph{ Distributed decryption.} 
Multiple parties hold $\Enc(x)$, 
and the decryption trying to distributedly decrypt it ($\Enc(x)$ is the form of BGV encryption), where each party has a piece of the secret key. 
} 

%% file: robust-and-verifiable-mpc.tex
\section{Robust and Verifiable MPC}\label{sec:protocol}
In this section, we first describe the relevant ideal functionalities and then describe how our protocol securely realizes these functionalities following the simulation paradigm (cf.~\cite{Canetti05}).
\subsection{Ideal Functionalities}
The ideal functionality for MPC is depicted in Fig.~\ref{fig:Fmpc}. Each party $P_i$ (note that input parties may be different from server $S_{i}\in$$\cal S$) provides its input $in_i$ for circuit $C$, then obtain output OUT $=C(in_{0},in_{1},...,in_{n-1})$.  
    \begin{figure}[htbp]
\funcbox{\Fmpc}{
\vspace{-.4in}
{\small
\begin{tiret}
   \item {\bf INIT:} On input $(\mathsf{init}, C_f, p)$ from all parties(where $C_f$ is a circuit with $n$ inputs and one output computing $f$, consisting of addition and multiplication gates over $Z_p$)
   \begin{newenum}
       \item Store $C_f$ and $p$
       \item Wait for ${\cal A}$ to provide the set ${\cal I}$ of adversarially controlled party indices
       \item Store OUT $:= \bot$
   \end{newenum}
\item {\bf INPUT:}
 On input $(\mathsf{input}, P_i, in_i)$,
 store (INPUT, $P_i$, $in_i$)
 \item {\bf EVAL: }
On input ($\mathsf{eval}$) from all parties:
\begin{newenum}
    \item If not all input values have been provided, output REJECT
    \item Evaluate the circuit $C_f$ on inputs ($in_1, ..., in_n$). When the evaluation is completed, store the resulting value as OUT
\end{newenum}
\item {\bf OUTPUT}: On input ($\mathsf{output}$) from all parties:
\begin{newenum}
    \item Send ($\mathsf{output\textit{-}result}, \textrm {OUT}$) to all parties $P_i$   
\end{newenum}
\end{tiret}
 \vspace{.01in}
} 
}
\vspace*{1mm}
\caption{
The ideal functionality for secure multi-party computation (MPC).}
\label{fig:Fmpc}
\end{figure}
Figure~\ref{fig:Fcidampc} depicts the ideal functionality for MPC with completely identifiable abort MPC ($\Fcidampc$); when malicious behavior is detected, the functionality will abort and output the identity of the misbehaving
party.
\begin{figure}[htbp]
\funcbox{\Fcidampc}{
\vspace{-.45in}
{\small
\begin{tiret}
   \item {\bf INIT:} Same as $\Fmpc$. 
   
In addition, 
   receive and record the identity of trusted client $C$.
 Set $L_\mathrm{cheat} \coloneqq \emptyset$

   \medskip
\item {\bf INPUT, EVAL:}
 Same as $\Fmpc$.

\medskip
\item {\bf OUTPUT}: On input ($\mathsf{output}$) from all parties:
\begin{newenum}
    \item Send ($\mathsf{output\textit{-}result}, \textrm{OUT}$) to all adversarially controlled parties $P_i \in$ ${\cal I}$.
    \item Run ABORT, waiting for each adversarially controlled party to send either ($\mathsf{abort}, \mathrm{ACCEPT}$) or ($\mathsf{abort}, \mathrm{ABORT}$).
    \item Send ($\mathsf{output\textit{-}result}, \textrm{OUT},  L_\mathrm{cheat}$) to all parties, where $\textrm{OUT}$ may now be $\bot$
\end{newenum}

\item {$\bf ABORT:$}
On input ($\mathsf{abort}, x_i$) from 
an adversarial server $S_i$
    \begin{newenum}
        \item $L_\mathrm{cheat} := L_\mathrm{cheat} \cup S_i$
        \item Set $\textrm{OUT}$  $:= \bot$
    \end{newenum}
\end{tiret}
}
} 
\caption{Ideal functionality for MPC with {\em completely identifiable abort}.}
\label{fig:Fcidampc}
\end{figure}

Combining the $\Fcidampc$ approach with a ``trusted dealer'' robustness, can be achieved for a number of corruptions of up to
$n-2$ parties.  
For the reason for dissimilar thresholds (i.e., $n-2$ vs. $n-1$), please refer to~\ref{subsec:contributions}.
The functionality $\Fcidarvmpc$ for robustness with public verifiability at Fig.~\ref{fig:Ideal function of Fcidarmpc}. 
\begin{figure}[htbp]
\funcbox{\Fcidarvmpc}{
\vspace{-.4in}
{\small
    \begin{tiret}
   \item {\bf INIT:} Same as $\Fcidampc$, additionally receive and record the identity of trusted client ${\cal T}$. Set $L_{cheat} = \emptyset$
\item {\bf INPUT:}
 Same as $\Fcidampc$
 \item {\bf EVAL: }
Same as $\Fcidampc$
\item {\bf OUTPUT}: Same as $\Fcidampc$
\item {$\bf ABORT:$}
On input ($\mathsf{abort}, x_i$) from
an adversarial server $S_i$
\begin{newenum}
        \item Add $S_i$ to $L_{cheat}$
        \item If $|L_{cheat}|\geq n-1$, set $\textrm{OUT}$ = $\bot$
\end{newenum}
\item {\bf AUDIT CESS:}
On input ($\mathsf{audit\textit{-}CESS}$) from ($\mathsf{audited\textit{-}CESS}$), outputs ($\mathsf{audited\textit{-}CESS}, L_{cheat}$)
\end{tiret}
} 
}
\caption{Ideal functionality for robust MPC with completely identifiable abort with public verifiability.}
\label{fig:Ideal function of Fcidarmpc}
\end{figure}
\ignore{\begin{figure}[htbp]
\funcbox{$\Fcidarvmpc$}{
\vspace{-.4in}
{\small
   \item {\bf INIT:} Same as $\Fcidarmpc$
\item {\bf INPUT:}
 Same as $\Fcidarmpc$
 \item {\bf EVAL: }
Same as $\Fcidarmpc$
\item {\bf OUTPUT}: Same as $\Fcidarmpc$
\item {\bf ABORT:}
Same as $\Fcidarmpc$
\item {\bf AUDIT CESS:}
On input ($\mathsf{audit\textit{-}CESS}$) from ($\mathsf{audited\textit{-}CESS}$), outputs ($\mathsf{audited\textit{-}CESS}, L_{cheat}$)
} 
}
\caption{\juan{Check:} Adding third-party verifiability to the MPC functionality.}
\label{fig:Ideal function of Fcidarvmpc}
\end{figure}
}
\subsection{Protocol Description}
At a high level, the protocol consists of an offline phase and an online phase.
In the offline phase, the client generates commitment and encryption parameters (including commitment's public parameters and encryption's public/private keys)
and hands them to the STTPs. Next, the client and STTP collaboratively generate input shares, along with the corresponding commitments and homomorphic encryptions. The objective is to ensure that the STTP does not possess all the input shares and their associated encryptions, but instead holds only the commitments to the input shares.
After that, shares, commitments, and encryptions are distributed to the corresponding server. We call our protocol $\Pi_{\textrm{RV-MPC}}$, which is split into two parts: offline and online.

In the online phase, the servers are responsible for carrying out the computation.
If a malicious behavior is detected by any of the servers, the server makes an accusation to STTP. STTP uses the commitment to validate the accusation; if the accusation is valid, STTP broadcasts the encryption's secret key of the accused server.
Next, all the servers use the received secret key to recover the malicious share held by the accused server.
We now turn to a more detailed specification of the protocol. 
\paragraph{Offline phase.}
As previously noted, the offline protocol is designed to allow the STTPs (In our settings, we have multiple offline phase STTPs and one online phase STTP) to compute the randomness utilized in the online phase, such as Beaver triples. In this phase, random elements $rs$ are generated to be used in masking inputs and distributing input shares accordingly.

To ensure that no single STTP can access the secret input, the offline protocol employs homomorphic encryption and distributed decryption schemes (see Section~2). 
This approach simplifies security by leveraging the semi-honest assumption for STTPs, as our setting does not require the additional complexity presented in \cite{DBLP:conf/sp/RiviniusR0K22}. 
Here, we also assume that the client remains honest, solely providing inputs and receiving outputs without further involvement. The structure of the offline phase is illustrated in Fig.~\ref{fig:offline-phase}.

\begin{figure}[htbp]

\protbox{\Pi^{\textit{off}}_{\textrm{RV-MPC}}}{
{\small
\begin{tiret}
    \item Parties set up BGV parameters (similarly to the distributed decryption protocol).
\item For each party $P_i$:
\begin{enumerate}
    \item Sample $W_i \leftarrow U(R_p^{(I+3m)*t})$ and $y_i \leftarrow U(R_p^{I+3m})$
\item Encrypt $W_i$ and $y_i$ to get $\Enc(W_i)$ and $\Enc(y_i)$, then
broadcast $\Enc(W_i)$ and $\Enc(y_i)$
\item Commit to $y_i$ with $R_c$($y_i$), gets $\Comm(y_i)$ then
broadcasts $\Comm(y_i)$
\item Compute $\Enc(W) = \Enc(W_i)$ 
\item For each $P_j \in P$:
\begin{enumerate}
    \item Define encrypted share of $v$ = $W[, 0]$ as $\Enc([v]_j) =  \sum_{l=0}^{t-1} j^{l}\Enc(W)[\cdot,l]$
    \item $m_j$ =Dist-Decrypt $(\Enc(y_i)-\Enc([v]_j))$
\end{enumerate}
\item Construct $\langle v_i \rangle = (y_i-m_i, R_c(y_i), \Comm(y_1)-m_1,…,\Comm(y_n)-m_n)$
\item Split $\langle v \rangle_i$ and $\Enc(v)$ in parts of size I, M, M, M to get views and ciphertexts for r, a, b, d.
\item For the Beaver triples:
\begin{enumerate}
\item Compute $\Enc(c)$ with $\Enc(a)*\Enc(b)$ (using the homomorphic property of encryption)
\item Compute $\langle c \rangle_i$ =  $\langle d \rangle_i$+Dist-Decrypt$(\Enc(c) - \Enc(d))$
\item Each party $P_i$ sends its share to $STTP_i$
\item STTPs then sends shares of $r, a, b, c$ to the client $\cal C$.
\end{enumerate}
\end{enumerate}
\end{tiret}
\vspace{.01in}
} 
}
\caption{The protocol's offline phase.}
\label{fig:offline-phase}
\end{figure}
\ignore{
\begin{figure}[htbp]
\protbox{\Pi_{\textrm{Input}}}{
\juan{Revise name.}
{\small
	Client $\cal C$ intends to distribute input $x$. The simplest way is for $\cal C$ to generate randomness $r$ in the offline protocol and compute secret shares. Then $\cal C$ chooses parameters for the homomorphic encryption, computes the homomorphic encryptions and lattice-based commitments, broadcasts the lattice-based commitments, sends the homomorphic encryptions to all the servers, sends each share to the corresponding server, and sends the decryption keys of the homomorphic encryption to the online STTP that monitors the online computation.

 \begin{newenum}
     \item $\cal C$ creates a share $x_j = x-r+r_j$, where $x$ = $\sum x_j$ and $1 \leq j \leq n$ ($n$ is the number of servers)
\item $\cal C$ computes and broadcasts $\Enc(sk_{j}, x_j)$
\item $\cal C$ computes and broadcasts $\Comm(x_j)$
\item $\cal C$ sends $sk_j$ to the online STTP
\end{newenum}

\vspace{.1in}
} 
}
\caption{The input protocol.}
\label{fig:input-proc}
\end{figure} 
}
 Here's an extension of $C$ offloading the computation work to STTPs. Suppose there are $n$ servers (denote each server as $S_i$) and $n$ STTPs (denote each STTP as $STTP_i$). The goal is to have each server $S_i$ hold share $x_i$ along with commitments and homomorphic encryptions for all shares.
 
 $\cal C$ sends the private homomorphic encryption keys to all STTPs, then distributes share $x_i$ to $STTP_i$. Each $STTP_i$ computes the homomorphic encryptions and lattice-based commitments of $x_i$, broadcasts the lattice-based commitments, sends the homomorphic encryptions to all the servers, and delivers each share to the corresponding server. Note that each $STTP_i$ only gets a single share $x_i$, which is random with respect to the real value $x$.

Another extension involves having the servers compute the offline phase rather than the STTPs. In this scenario, it is necessary to prevent servers from behaving maliciously without getting identified. To address this, the client can generate a secret and private key pair for each party. The client then broadcasts the public key and sends each secret key to the respective parties and the STTP. Once the secret and public keys have been distributed, the parties can execute the offline protocol as described in \cite{DBLP:conf/sp/RiviniusR0K22}.
\paragraph{Online phase.}
The online protocol comprises a set of servers ${\cal S}$ carrying out a computation without leaking the input to any of them.
 {\em CESS} stands for {\em commitment-enhanced secret sharing}, and was introduced in ~\cite{DBLP:conf/icits/CunninghamFY17}. Its objective is to realize, in dishonest majority settings, 
{\em completely identifiable abort},
meaning that {\em all} parties that misbehave are flagged as malicious. In order to make a CESS-type protocol practical,  Rivinius {\em et al.}~\cite{DBLP:conf/sp/RiviniusR0K22} proposed a lattice-based commitment scheme, which only takes approximately 20x time as MP-SPDZ \cite{DBLP:conf/crypto/DamgardPSZ12} when there are 2 servers, whereas the approach in
\cite{DBLP:conf/icits/CunninghamFY17}
using Pedersen commitment would be roughly 800x. 
The protocol in \cite{DBLP:conf/sp/RiviniusR0K22} realizes the completely identifiable approach, and combining \cite{DBLP:conf/sp/RiviniusR0K22} with STTP, robustness can be achieved for a number of corruptions of up to $n-2$ parties.  
For the reason for dissimilar thresholds (i.e., $n-2$ vs. $n-1$) please refer to \ref{subsec:contributions}
The input secret sharing phase allows the client to secret share its input and broadcasts the commitment and encryption of all inputs to the computation parties.

The robust protocol's precondition is as follows: For each client input $x$, each server $S_i$ holds  $([x]_{i}, r_{i}$, \Comm($x_{1}$),$\dots$,\Comm($x_n$), \Enc($x_1$),$\dots$, \Enc($x_n$)). 
The semi-trusted third party (STTP) holds all commitments. Additionally, each $S_i$ holds the Beaver triples received from the offline phase as well as the commitments of all Beaver triple shares. 
We describe our protocol, which achieves robustness up to $n-2$ malicious parties in Fig.~\ref{fig:robust-protocol}. Our protocol achieves a stronger security property than protocols in \cite{DBLP:conf/sp/RiviniusR0K22,DBLP:conf/icits/CunninghamFY17}, which only achieve a completely identifiable abort in a dishonest majority setting.

\begin{figure}[htbp]
\protbox{\Pi^{\textit{on}}_{\textrm{RV-MPC}}}{

\vspace{-.15in}
{\small
\begin{tiret}
\item {\bf Input secret sharing:}
Client $\cal C$ intends to distribute input $x$. The simplest way is for $\cal C$ to generate randomness $r$ in the offline protocol and compute secret shares. Then $\cal C$ chooses parameters for the homomorphic encryption, computes the homomorphic encryptions and lattice-based commitments, broadcasts the lattice-based commitments, sends the homomorphic encryptions to all the servers, sends each share to the corresponding server, and sends the decryption keys of the homomorphic encryption to the online STTP that monitors the online computation.

 \begin{newenum}
     \item $\cal C$ creates a share $x_1j = x-r+r_1j$, and $x_k=r_k$ where $rx$ = $\sum rx_j$ and $1 \leq j \leq n, 2\leq k \leq n$ ($n$ is the number of servers)
\item $\cal C$ computes and broadcasts $\Enc(sk_{j}, x_j)$
\item $\cal C$ computes and broadcasts $\Comm(x_j)$
\item $\cal C$ sends $sk_j$ to the online STTP
\end{newenum}

\vspace{.1in}
\item {\bf Preconditions:}
\begin{tiret}
\item Let $x_j$ denote the
share held by server $\mathcal{S}_j$
\item All servers and the 
STTP hold
$\Comm(x_i)$, $1 \leq i \leq n$.
\item All servers hold 
$\Enc(x_i)$ and $\Enc(r_i)$, $1\leq i \leq n$. 

\end{tiret}
\medskip
\item {\bf Online computation:}
\begin{newenum}
    
\item All servers update $\Enc(x_i)$ 
 and $\Enc(r_i)$ as the computation proceeds; denote 
 the updated ciphertexts as $\Enc(x^{'}_i), \Enc(r^{'}_i)$.
\item When $\mathcal{S}_k$ is identified as malicious,
STTP broadcasts $\mathcal{S}_k$'s decryption key.
The other servers decrypt $\Enc(x^{'}_k), \Enc(r^{'}_k)$, and obtain and $x'_k$, $r’_k$.
\item To recover the computation from failure, a designated server (e.g. $\mathcal{S}_0$), adds $x_k$ to its share (e.g., $x'_1 = x_1+x_k$). Since $x_k$ is a now a constant, all parties can locally update $\Comm(x_1)$ and $\Enc(x_1)$ by the homomorphic property of the commitment and encryption schemes. 
\item All servers send all $x'_{k}$ and $r'_k$ of the malicious server to STTP. Denote by ($x'_{k}$, $r'_{k}$) sent from server $S_i$ as ($x'_{ki}$, $r'_{ki}$). 
\item STTP then checks if there exists an inconsistency between all ($x'_{ki}$, $r'_{ki}$). If there is no inconsistency, accept $x'_{ki}$, $r'_{ki}$), and update the commitment. 
\item Else, do as follows: 
\begin{enumerate}

\item Set $M \leftarrow \emptyset$
\item While ($\exists$ ($x^{'}_{kj}$, $r^{'}_{kj}$)!= ($x^{'}_{ki}$, $r^{'}_{ki}$):
\begin{enumerate}
	\item Check the specific ($x^{'}_{kj}$, $r^{'}_{kj}$) and ($x^{'}_{ki}$, $r^{'}_{ki}$) \footnote{Happens at most $n-2$ times, since each check eliminates at least one party.}
	\item Identify the malicious pair  ($x^{'}_{km}$, $r^{'}_{km}$) with the commitment.
	\item Ignore the malicious pair  ($x^{'}_{km}$, $r^{'}_{km}$) and update $M := M \cup m$ 
\item Accepts ($x^{'}_{ki}$, $r^{'}_{ki}$) that remain honest, and update the commitments.
\end{enumerate}
\item For those m$\in$M, go back to step 3
\end{enumerate}
\end{newenum}
\end{tiret}

} 
}
\caption{Our robust and publicly verifiable MPC protocol (online phase).}
\label{fig:robust-protocol}
\end{figure}

Additionally, we describe the optimized commitment opening protocol in \cite{DBLP:conf/sp/RiviniusR0K22} here.
When opening a commitment, it is intuitive to just decommit (directly send the committed message and randomness that generates the commitment). However, directly decommit will require the commitment scheme to be equivocal to prove simulated secure \cite{DBLP:conf/icits/CunninghamFY17}.

The equivocation property enabled the simulator to open to any message (e.g., open $\Comm(x)$ to $x^{'}$, where $x^{'}\neq x$) but led to larger parameters and worse efficiency.
To get rid of the necessity of equivocation properties of the commitment scheme, \cite{DBLP:conf/sp/RiviniusR0K22} introduced a new commitment open protocol. The sender makes a new commitment, 
committing to the same message as the original commitment. Then the sender proves in zero-knowledge that the new and original commitment commit to the same message.
The 
opening protocol in \cite{DBLP:conf/sp/RiviniusR0K22} only requires a programmable RO
instead of an equivocal commitment.
As a result, without the equivocation property, the parameters of the commitment can be much smaller, leading to improved efficiency. For more details, please refer to \cite{DBLP:conf/sp/RiviniusR0K22}.

\subsection{Security Proof}
In this section, we 
argue 
the security 
of protocol
$\IVMPC$. 
\begin{theorem}  $\IVMPC$ realizes 
$\Fcidarvmpc$ in the ($\func{_\mathrm{PKI}},
\func{_{\mathrm{CRS}}}$)-hybrid model.
\end{theorem}

In \cite{DBLP:conf/crypto/DamgardPSZ12}, the authors prove that $\Pi_{SPDZ}$ realizes $\Fmpc$; further, in~\cite{DBLP:conf/icits/CunninghamFY17}, it is proven
that $\Pi_{CESS}$ realizes $\Fcidampc$.
This subsection aims to prove $\IVMPC$ realizes $\Fcidarvmpc$. One can observe that the additional feature $\IVMPC$ does compared to $\Pi_{CESS}$ is to provide public verifiability and robustly open some malicious shares. For public verifiability, credited from \cite{DBLP:conf/icits/CunninghamFY17}, since all exchanged messages are public, the simulation is trivial. As for robustly open malicious share, the simulation could be achieved by having the simulator choose the public and private keys of all servers in ${\cal S}$, then generate/open the homomorphic encryptions. Given the indistinguishable property of homomorphic encryption under different keys, one can not distinguish the homomorphic encryptions generated by the simulator from the homomorphic encryptions in the real world.

Another difference is that our scheme is a client-to-server mode (the client sends the inputs to the server), \cite{DBLP:conf/sp/RiviniusR0K22,DBLP:conf/crypto/DamgardPSZ12,DBLP:conf/icits/CunninghamFY17} are in a pure server mode.
To build a transformation from pure server mode to client and server mode, we observe the following:
In the pure server mode, we have two types of input secret sharing: malicious server input secret sharing and honest server input secret sharing.
In our case, we are in the model where the client provides the input shares. As we assume the client is always honest, we could view our model in the same way as an honest server input secret sharing.\\
\proofsketch{
Our proof is a combination of techniques used in  \cite{DBLP:conf/crypto/DamgardPSZ12,DBLP:conf/icits/CunninghamFY17,DBLP:conf/sp/RiviniusR0K22}.
For the simulator to simulate, it initially provides dummy input shares that add up to zero. After getting back the output from the ideal function, the simulator adjusts the output shares held by the honest parties to be consistent with the output. For example, the simulator has output $y^{'}$ that is computed using dummy input shares, and $y$ that is returned by the ideal function, the simulator picks one of the honest parties and then adds $y-y^{'}$ to the share held by the selected input party, so now the output will also be $y$ instead of $y^{'}$ (c.f. Appendix A.3 of \cite{DBLP:conf/crypto/DamgardPSZ12}). 
For the homomorphic encryption, we do not need to adjust it as discussed above. For opening the commitments, as discussed in \cite{DBLP:conf/sp/RiviniusR0K22,DBLP:journals/jacm/LyubashevskyPR13}, with a programable random oracle, the simulator can fake a zero-knowledge proof for the final open, which is indistinguishable from the real protocol.

The simulator 
for $\IVMPC$ is shown in Fig.~\ref{fig:simulator-proof-protocol}.
} \hfill \QED

\begin{figure}[h!]
\simbox{S_\textrm{RV-MPC}}{

\vspace{-.15in}

{\small
\begin{tiret}
\item {\bf INIT:}
$S_\textrm{RV-MPC}$ provides a CRS, such that it knows the lattice-based commitment parameters. The simulator then chooses the public/private key of the homomorphic encryption used for robustness.\\
Then $S_\mathrm{\IVMPC}$ distributes shares, commitments, and homomorphic encryptions to each server as below:
\item {\bf INPUT:} For each input, $S_\textrm{RV-MPC}$ generates dummy shares of 0, generates the commitments and homomorphic encryptions, and then distributes them to each server.
\item {\bf EVAL:} $S_\textrm{RV-MPC}$ evaluates circuit C gate by gate

\item {\bf OUTPUT:} $S_\textrm{RV-MPC}$ gets the output from functionality $\func{_\text{RV-MPC}}$\\ then the simulator $S_\textrm{RV-MPC}$ proceeds as follows:
\begin{enumerate}
    \item If the $L_{cheat} >=n-2$, the simulator aborts.
    \item Else, modify the output share of one of the honest servers to be consistent with the output $out$. (Suppose the output from $\func{_\textrm{RV-MPC}}$ is $out$, and with the dummy 0 share, we get $out^{'}$, add $out-out^{'}$ to an output share of one of the honest servers)
    \item For the zero-knowledge proof for the final output, as stated in \cite{DBLP:conf/sp/RiviniusR0K22}, the simulator can fake a ZKP with the programmable random oracle.
\end{enumerate}
\item {\bf OPEN:} The simulator is able to do so since it holds the public and private keys for all homomorphic encryptions. The simulator sends the decryption key to the servers when it detects a server that is malicious. Furthermore, the dishonest parties' shares are distributed uniformly, therefore, the open view between the real and ideal world will be indistinguishable.

\item {$\bf AUDIT:$} The simulator can do an audit since the commitments can be opened to the public (with hiding property, the commitment does not leak info about the share).
\end{tiret}
\vspace{.05in}
} 
}
\caption{Simulator for $\Pi_\textrm{RV-MPC}$'s proof.}
\label{fig:simulator-proof-protocol}
\end{figure}

%% file: applications-and-experiment-results.tex
\section{Applications and Experimental Results}
\label{sec:experiments}
\subsection{Network-A Benchmark}
\label{subsec:network-A}
In this section we first benchmark Network-A~\cite{DBLP:conf/sp/MohasselZ17,DBLP:conf/ccs/Keller20,DBLP:conf/sp/RiviniusR0K22} with our protocol. 
For the environment, we have three computation servers, where up to two can be malicious. In addition, we set up an STTP party to provide robustness.
We use the same parameters for lattice cryptography primitives of \cite{DBLP:conf/sp/RiviniusR0K22}, where the parameter of computation security is 40 bits.
Furthermore, we calculate the size of the homomorphic encryption we used with our robust approach by \cite{DBLP:journals/jmc/AlbrechtPS15}, we have BGV encryption with 350 bits. 

We ran our experiments on machines with 32GB RAM and 16 vCPUs.
Below is the benchmark of our computation online run time (in seconds), compared to the SPDZ 
and \cite{DBLP:conf/sp/RiviniusR0K22} protocols. We observe that,
compares to SPDZ, the running time of our protocol is about 65x (see table~\ref{tab:efficiency_comparison}).


\begin{table}[h!]
\caption{Comparison of efficiency of different protocols, benchmarking on network-A. Columns 2 and 3 represent amortized computation times.}
\centering
\begin{tabular}{|c|c|c|}
  \hline
  SPDZ (LowGear) 
  & \cite{DBLP:conf/sp/RiviniusR0K22} 
  & Our protocol  \\
  \hline
   $\approx$ 0.0036s
   & $\approx$ 0.135s
   & $\approx$ 0.227s \\
  \hline
\end{tabular}

\label{tab:efficiency_comparison}
\end{table}

Furthermore,  we also show that our protocol recovers quickly when a malicious server is detected (see table~\ref{tab:recovery_comparison}). Since the malicious server will be eliminated from the computation, the recovery time can be offset by having one less server in the computation. 
In the experiment, we show the 
time to recover the 
share from the malicious party, plus the time the computation continues with the two remaining parties is not much different compared to the three-party computation 
when no malicious behavior is detected. 

\begin{table}[]
\caption{The three columns from left to right indicate: time of recovery from malicious shares, recovery time plus remaining computation with two parties, and run time if three parties behave honestly (All in amortized measurement).}
\label{tab:recovery_comparison}
\resizebox{\columnwidth}{!}{%
\begin{tabular}{|c|c|c|}
\hline
Recovery time & \begin{tabular}[c]{@{}c@{}}Recovery time + \\ remaining computation \\ with two parties\end{tabular} & \begin{tabular}[c]{@{}c@{}}Our protocol \\ with 3 parties\end{tabular} \\ \hline
$\approx$ 0.096s  & $\approx$ 0.211s                                                                                         & $\approx$ 0.227s                                                           \\ \hline
\end{tabular}%
}
\end{table}

\input{evaluation-framework}

%% file: evaluation-framework.tex
\subsection{ML Inference Framework}
\label{subsec:ml-framework-design}
In this subsection, we first present the design of a framework for privacy-preserving machine learning (ML) inference, employing MPC protocols under malicious-dishonest majority security settings, followed by the evaluation of the lattice-based MPC protocol proposed in our study.

\subsection{Framework design.}
\begin{figure}[h!]
  \centering
  \includegraphics[scale=0.40]
  {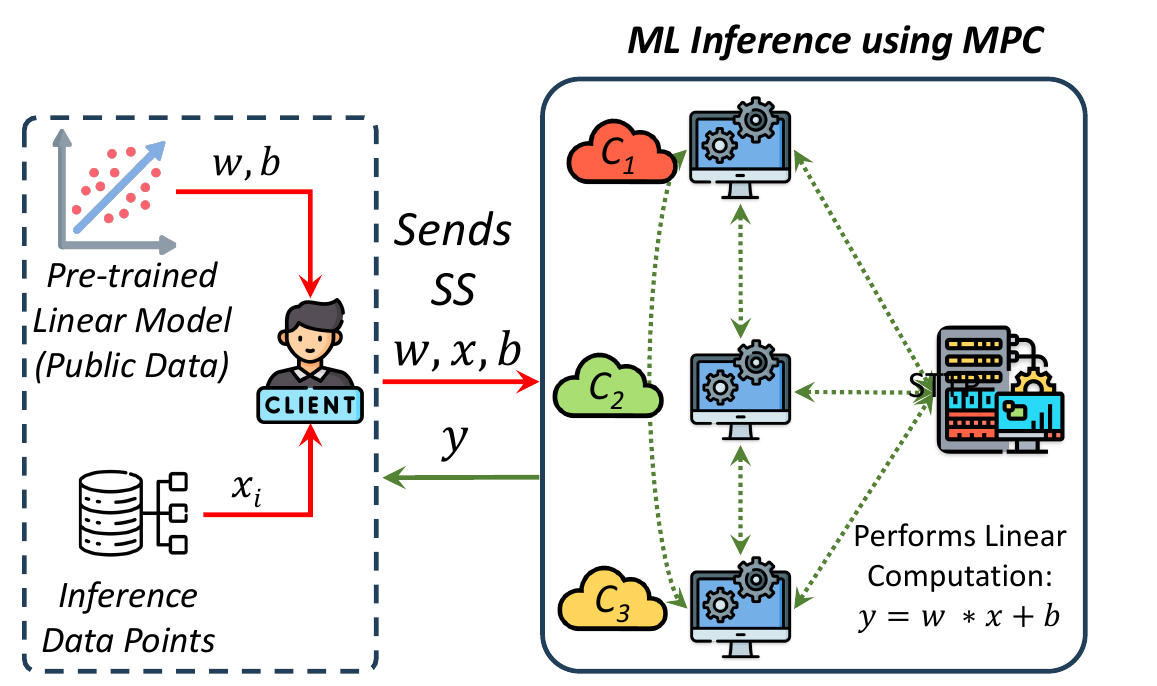}
  \caption{A Design of ML Inference Framework}
  \label{fig:system-design}
\end{figure}

Our framework enables secure inference using a pre-trained linear model, while ensuring the confidentiality of both the model and the inference input data. The overall framework design is depicted in Figure~\ref{fig:system-design}.

The linear model is trained on publicly available data, and model parameters comprising weights (\texttt{w}) and biases (\texttt{b}), which are subsequently secret-shared among computation parties involved in MPC protocol. Similarly, inference input data point(s) ($x_{i}$) are transformed into secret shares to safeguard user privacy. These secret shares are shared and distributed among computational parties, ensuring that no single computation party has access to the original data or model parameters.

The client initiates the process by sending secret shares of the data, which are to be processed, along with the secret shares of the weights and biases to the MPC servers. These servers perform linear computations, specifically computing $\texttt{w} \cdot \texttt{x} + \texttt{b}$, where w represents the weights, \texttt{x} represents the input data, and \texttt{b} represents the bias. This computation is performed on encrypted secret shares, ensuring the privacy of the data.

Once the MPC servers have completed the necessary computations, the results are securely transmitted back to the client in encrypted form. Upon receipt, the client decrypts these results to proceed with further data processing specific to the model used. This includes the application of activation functions and thresholding to finalize the inference process. For instances utilizing the Logistic Regression model, the decrypted output is first processed through a logistic function to map the computed values to probabilities, followed by a thresholding step to categorize these probabilities into discrete class labels.



\subsection{Framework evaluation.}
In this section, we define the evaluation framework to evaluate the Lattice-based MPC protocol proposed in our study. We describe the datasets, experiment settings, and metrics for validating the correctness and efficiency of the proposed MPC protocol in performing ML inferences. Additionally, we compare the performance of our Lattice-based protocol with MASCOT \cite{MASCOT-MKOESP-16}, MASCOT* \footnote{MASCOT* refers to the MASCOT protocol configured with multiple MACs to enhance the security parameter, making it a multiple of the prime length \cite{DBLP:conf/ccs/Keller20}.}, SPDZ2k \cite{SPDZ2k-RCIDDEOS-18}, and LowGear \cite{LowGear-MKVPDR-18} MPC protocols, all evaluated under the Malicious Dishonest Majority security setting.

\paragraph{Description of the datasets} In this study, we assessed the performance of the proposed MPC protocol on the Wisconsin Breast Cancer dataset and a subset of the Iris flower dataset. 

The Wisconsin Breast Cancer dataset, made available by Wolberg {\em et al.} \cite{misc_breast_cancer_wisconsin_diagnostic_17}, is widely used in ML research and includes features derived from digitized images of fine needle aspirates (FNA) of breast masses. These masses are categorized into two classes: benign or malignant. The dataset contains 569 instances, each with 30 attributes or features. These gestures describe various metrics for the tumors: radius, texture, perimeter, area, smoothness, compactness, concavity, concave points, symmetry, and fractal dimension. This dataset is frequently utilized in ML research making it a benchmark for comparing the performance of various classification algorithms for binary classification tasks. 


The Iris flower dataset, introduced by Fischer \cite{iris-dataset}, is a classical dataset in ML research. It contains 150 samples, distributed across three classes of Iris species: Iris-setosa, Iris-versicolor, and Iris-virginica. Each instance is described by four features: sepal length, sepal width, petal length, and petal width. Due to its simplicity and balanced structure, the dataset is commonly used for classification tasks, particularly for exploring the performance of algorithms in multiclass classification scenarios. 

For our study, we focused on a subset of the Iris dataset, considering only the Iris-setosa and Iris-versicolor classes to formulate a binary classification task. From the 100 relevant samples (50 samples per class), we used 30 samples from each class for training and reserved 20 samples from each class for testing. This setup ensured a balanced dataset for training while providing an unseen test set for evaluating the performance of the classification model. This selection of balanced number of samples for the samples from each class ensures that the classifier is trained on a balanced data set from both classes, thereby avoiding bias for a particular class and allowing robustness evaluation on unseen data.

\paragraph{Experiment settings} In this study, we conducted experiments to evaluate the performance of a lattice-based MPC protocol, utilizing linear ML classifier, Logistic Regression. Our experiments were performed on Amazon Web Services (AWS) Cloud Virtual Machines (VMs) under two configurations: first, with all VMs situated within the same Cloud Service Provider (CSP) to ensure uniform computational resources and network conditions; second, with each computational party hosted on different CSPs to simulate a distributed environment with varying network conditions.


The Logistic Regression model was trained using the `ml` package available in MP-SPDZ. Following the training phase, the weights and biases of these models are extracted for evaluating lattice-based MPC protocol. More specifically, we assess the ability of the MPC protocol to perform secure and efficient computations. 

\paragraph{Assessing computation correctness} To validate the correctness of computations performed by the lattice-based MPC protocol, we used Accuracy as the evaluation metric, which measures the proportion of correctly predicted instances out of the total evaluated.

In our experiments, we compared the accuracy achieved in centralized settings (evaluating plaintext data directly) with that obtained using the MPC protocol. This comparison confirmed the correctness of computations under MPC and highlighted any potential efficiency losses due to its distributed nature.

The accuracy achieved with the MPC protocol (88.33\% for the Wisconsin Breast Cancer dataset and 100\% for the Iris Flower dataset) matched the centralized settings. This demonstrates that our protocol performs computations correctly, maintaining high precision comparable to traditional centralized methods while ensuring secure computation.




\paragraph{Comparative analysis of MPC protocols}
We conducted a comparative analysis of MPC protocols, focusing on three key metrics: inference times, size of data exchange, and number of communication rounds. The detailed analysis for each metric is presented below:

\begin{enumerate}[leftmargin=0.5cm]
    \item \textbf{Inference Time:} The inference time is defined as time required to compute an output from a trained model under an MPC setup. This performance metric is essential for assessing the efficiency of MPC protocols as it influences its scalability in privacy-preserving application.

     \textbf{All VMs hosted on same CSP:} Our analysis of inference times for various Multi-Party Computation (MPC) protocols across the Iris and Breast Cancer datasets reveals no clear pattern in performance superiority except in the case of Lattice-based protocol. The MACOT, MASCOT* (mama), SPDZ2k, and LowGear protocols display closely competitive inference times on both datasets.
    
    For the Iris Dataset, the inference times are remarkably close, with the fastest (MASCOT at 0.0026122 seconds) and the slowest (MASCOT* at 0.0028984 seconds) among them differing by less than 0.0003 seconds. A similar trend is observable in the Breast Cancer Dataset, where the range between the fastest (MASCOT at 0.00472605 seconds) and the slowest among the traditional protocols (MASCOT* at 0.00506337 seconds) remains narrow.
    
    
    Conversely, the Lattice-based protocol records the highest inference time at 0.0771 seconds on Wisconsin Breast Cancer Dataset and 0.0143 seconds on Iris Flower Dataset. While this may seem less performance-driven, the Lattice-based protocol offers several advantages in malicious scenarios. Unlike other protocols, Lattice-based protocol is designed to continue computations without restart even if a malicious behavior is detected, thus maintaining the integrity of the computation process without the need for time-consuming restarts. This ensures operational continuity and is not available in other MPC protocols. Additionally, Lattice incorporates mechanisms to identify and handle cheaters effectively, ensuring that the computation completes successfully. While other protocols cannot identify cheaters leading potentially to indefinite computation loops. This robustness assurance that computations will complete regardless of adversarial behavior within the protocol participants provides a stronger argument for its use.

    \textbf{All VMs on different CSPs:} To simulate the scenario where each computation party is located on different Cloud Service Providers (CSPs), we utilized virtual machines (VMs) on the same cloud service but deployed them in different geographic locations. Specifically, we selected three AWS regions: N. Virginia, N. California, and Ohio. Each of the three computation parties was hosted in one of these regions. This configuration emulates a multi-cloud environment by introducing network variability and latency similar to what would be experienced if the parties were on different CSPs. 

    Our results on Iris dataset show that SPDZ2k achieved lowest inference time at 0.36095 seconds. The other protocols—MASCOT, MASCOT*, and LowGear—exhibited slightly higher inference times, ranging from 0.36145 to 0.36160 seconds. The minimal differences suggest that these protocols have comparable computational overheads in a distributed cloud environment with geographically dispersed VMs. The lattice protocol only runs for 0.0176 seconds. This is an example of how amortizing many instances leads to better utilization of hardware resources. Lots of instances will fill the network buffer/blocks, therefore, when considering network latency, it does not suffer so much. Furthermore, batching multiple instances also makes it more computation-intensive, and suffers less when considering network latency. 

    Similarly, for the Breast Cancer dataset, SPDZ2k demonstrated the lowest inference time at approximately 0.41599 seconds. The inference times for MASCOT, MASCOT*, and LowGear were slightly higher, between 0.41660 and 0.41675 seconds. This consistency across datasets reinforces the efficiency of SPDZ2k in handling distributed computations over disparate geographic regions. The lattice protocol only runs for 0.1107 seconds. This also shows that the lattice protocol is more robust when introducing network latency.
    
    \item \textbf{Data Exchange:} The data exchange refers to the amount of data that needs to be communicated or exchanged between the computation parties during the execution of the protocol. This could include sharing of secret-shares, exchanging encrypted data, or transferring computed values.
    
    For the Iris Flower Dataset, each party in the Lattice-based protocol sends 0.223 MB of data per party, resulting in global data exchange of just 0.669 MB. This significantly contrasts with the other protocols, where the global data sent ranges from 0.021928 to 0.022576 MB. Similarly, for the Wisconsin Breast Cancer Dataset, each party in the Lattice-based protocol sends 2.25 MB of data, resulting in global data exchange of just 6.75 MB. For other MPC protocols, the global data exchange for MASCOT, MASCOT*, and LowGear protocols is 0.309856 MB, and 0.31048 MB for the the SPDZ2k protocol.
    

    \item \textbf{Number of Rounds:} The "number of rounds" refers to the number of sequential communication steps required between the parties involved in the computation to complete a given MPC protocol. 
    
    Our results show that MASCOT, MASCOT*(mama), SPDZ2k, LowGear, and Lattice-based protocols shows varying number of rounds distributed across three computation parties. MASCOT, MASCOT*(mama), LowGear, and spdz2k show consistent pattern where party 1 engages in more communication rounds (21 rounds for Breast Cancer Dataset, and 17 rounds for Iris Dataset) compared to parties 2 and 3 (15 rounds for Breast Cancer Dataset, and 13 rounds for Iris Dataset each). This higher number of rounds for party 1 is because the party 1 is act as coordination server as well which distributes the data to other parties for computation and also accumulates the results from other parties once the computation is completed. Lattice-based protocol requires significantly more rounds (1860 rounds for the Wisconsin Breast Cancer Dataset and 200 rounds for Iris Flower Dataset for each party), but this is offset by its unique ability to identify cheating, ensuring that computations always complete successfully. This property of Lattice-based protocol guaranteeing completion of the computation despite presence of malicious behavior, makes is ideal choice for scenarios demanding high security and reliability.



\end{enumerate}

It is important to note that when we conducted the experiments under both configurations—computation parties located in the same region and those in different regions—the data exchange and number of rounds remained unchanged across all protocols. This consistency is expected, as the protocols' communication patterns and computational steps are predefined and independent of the physical locations of the computation parties. The only metric that exhibited variation was the inference time, which increased when computation parties were located in different regions due to the added network latency inherent in cross-regional communication. This observation underscores that while the efficiency of the protocols in terms of data exchanged and rounds required remains unaffected by geographic distribution, the actual performance time is influenced by the network conditions between the participating parties.

\subsection{Discussion} The variations in performance observed among the different MPC protocols in our study can be largely attributed to their underlying cryptographic mechanisms and computational models. While MASCOT, SPDZ2k and LowGear differ significantly in their preprocessing approaches, it is important to recognize that these protocols have an identical online phase. This similarity allows for a streamlined analysis of performance variations, as the primary differences originate from their distinct preprocessing states. 

LowGear introduces significant enhancements by optimizing the preprocessing phase with semi-homomorphic encryption (SHE), specifically the BGV scheme, which claims to be faster than MASCOT in both LAN and WAN settings by reducing communication and computational load \cite{LowGear-MKVPDR-18}. These improvements in the preprocessing phase enable the online phase to operate more efficiently, leveraging the precomputed data more effectively.

The choice between MASCOT, SPDZ2k, and LowGear might involve a trade-off between preprocessing efficiency and the potential benefits of a specific computational model during the online phase. While MASCOT and LowGear typically demonstrate better communication efficiency in preprocessing compared to SPDZ2k, the latter could offer advantages in certain online computations, particularly those involving comparisons or bitwise operations. This highlights how SPDZ2k’s approach to modulo $2^{k}$ computations, which aligns closely with standard CPU architectures, might be particularly beneficial for operations common in many practical applications.

The lattice-based protocol exhibits slightly lower efficiency compared to SPDZ protocols under normal conditions. However, in scenarios involving network latency, it demonstrates greater stability due to its amortized characteristics. We conclude that the amortized design of the lattice-based protocol allows for better hardware resource utilization, making it more robust across varying hardware and network environments. Combined with its enhanced security properties, this suggests that the lattice-based protocol holds certain advantages over SPDZ protocols.
